\documentstyle[preprint,aps]{revtex}   
\begin{document}   
\tightenlines
\draft   
\title{Summing up the perturbation series in the Schwinger Model}   
\author{Tomasz Rado\.zycki\thanks{Electronic mail:  
torado@fuw.edu.pl}}   
\address{Physics Department, Warsaw University, ul. Ho\.za 69, 00-681 
Warsaw, Poland}   
\date{\today}   
\maketitle   
\begin{abstract}
Perturbation series for the electron propagator in the Schwinger Model is
summed up in a direct way by adding contributions coming from 
individual Feynman diagrams. The calculation shows the complete agreement
between nonperturbative and perturbative approaches.
\end{abstract}   
  
A long-standing question in quantum field theory is the connection between
perturbation series and exact, nonperturbative results. It dates back to
the Dyson's paper~\cite{dys} in which the author, considering stability of
a system conditions, suggested that physical quantities and Green's
functions should be nonanalytic in the coupling constant $g$ around $g=0$.
This in turn should result in the divergence, usually factorial type, of
the perturbation series. This conjecture was supported by simple
models~\cite{calog63} among which the most widely considered was the
anharmonic oscillator and its field-theoretical counterpart --- the
$\phi^4$ theory~\cite{lang,graf,bw71,shirk77,lipat,brez,ng,zinn,bachas} as
well as by other, more realistic, field theories as QED for
instance~\cite{par,bog} (see~\cite{fischer} for further references).  In
these cases the required estimations for the nonperturbative results were
often obtained with the use of the generalized (Pad\'e, Borel) summation
methods (for a review of this approach see~\cite{fischer,kazak,jzj}).
There have also been found counterexamples, regarding the Dyson's
observation, in which the perturbation series is not divergent in spite of
instability (although it may be convergent to an incorrect
result)~\cite{herbst,calog79,beh}.
 
In QED the nonanalyticity in the coupling constant often manifests itself
through the presence of a logarithmic function of the fine structure
constant $\alpha$ in the calculated quantities~\cite{jauch,ibbtr,fett} and
in consequence means the divergence of coefficients in the Taylor
expansion in $\alpha$ (in other words divergence of Feynman diagrams)
resulting in necessity of infinite renormalisation. One can say that this
means the incorrectness of the perturbation
expansion~\cite{jbw1,jbw2,jb1,ad,jb2,ibb0,ibb,ibb1}.
 
Although the summability of the perturbation series still remains an
opened question, perturbation theory constitutes, however, the main tool
in practical calculations giving, especially in Quantum Electrodynamics,
excellent results. It seems, therefore, valuable to sum up directly the
perturbation series, by adding contributions of the individual Feynman
graphs, in a model theory in which the nonperturbative result is well
known. In this work we will concentrate on the 1+1 dimensional massless
QED known as the Schwinger Model~\cite{schw}. Up to our knowledge no such
direct summation has, in this model, been performed.  The focus will be
put on the electron propagator for which the explicit nonperturbative
formula in coordinate space was found~\cite{schw} (up to the final
$p$-integration) 
\begin{equation}  
S(x)={\cal S}_0(x)\exp\left[-ie^2\beta(x)\right]\; ,  
\label{propbet} 
\end{equation} 
${\cal S}_0$ being the free propagator. Function $\beta$ is defined by  
\begin{equation}  
\beta(x)=\left\{\begin{array}{ll}\frac{i}{2e^2}\left[- 
\frac{i\pi}{2}+\gamma_E +\ln\sqrt{ e^2x^2/4\pi}+   
\frac{i\pi}{2}H_0^{(1)}(\sqrt{e^2x^2/\pi})\right] & 
\hspace*{3ex}  
x\;\;\;\; {\rm timelike}\\   
\frac{i}{2e^2}\left[\gamma_E+\ln\sqrt{- 
e^2x^2/4\pi}+K_0(\sqrt{-e^2x^2/\pi})  
\right] &  \hspace*{3ex}x\;\;\;\; {\rm  
spacelike}\end{array}\right.\; . 
\label{beta}  
\end{equation}  
Symbol $\gamma_E$ denotes here the Euler constant and functions
$H_0^{(1)}$ and $K_0$ are Hankel function of the first kind, and Basset
function respectively~\cite{old}. 

One can expect that in this case the perturbation series should be
convergent and give, as the sum, the correct result since: 
\begin{enumerate}
\item no infinite renormalisation  has to be performed in the model 
\item if one reverses the sign of $e^2$, as suggested by Dyson, no collapse
should arise since in two dimensions the potential between equal sign
charges would be bounded from below
\item the appearance of a logarithm in equation~(\ref{beta}) is only 
apparent as the Hankel and Basset functions for small arguments --- which
means small values of the coupling constant (or small distances which is
equivalent here as the scale in the theory is imposed by $e$)--- behave
like 
\begin{equation} 
H_0^{(1)}(z)\approx \frac{2i}{\pi}(\ln x/2 +\gamma_E) +1 +  
{\rm analytic\;\; terms}\; , 
\label{asymp} 
\end{equation} 
and similarly for the $K_0$ function
\begin{equation}
K_0(z)\approx -\ln x/2 -\gamma_E + {\rm analytic\;\; terms}\; ,
\label{k0as}
\end{equation}
and nonanalytic functions cancel each other. The full propagator turns out
to be the free one in this limit (which corresponds also to the UV limit).
\end{enumerate}

The Schwinger Model may be characterized by the Lagrangian  
density  
\begin{equation}   
{\cal L}(x)=\overline{\Psi}(x)\left[i\gamma^{\mu}\partial_{\mu} -  
eA^{\mu}(x)\gamma_{\mu}\right]\Psi (x)-  
\frac{1}{4}F^{\mu\nu}(x)F_{\mu\nu}(x)-  
\frac{\lambda}{2}\left[\partial_{\mu}A^{\mu}(x)\right]^2\; .  
\label{lagr}   
\end{equation}   
The parameter $\lambda$ is here a gauge fixing one, and later will be  
set to infinity corresponding to the choice of the Landau gauge.  
  
In order to sum the perturbation series for the electron propagator one
first has to perform a presummation of vacuum polarization diagrams. It is
well known that this presummation is trivial since in this simple model
fermion loops with more than two vertices do not contribute and only
diagrams of Figure~\ref{vac} should be taken into account. It may be
easily checked by an explicit calculation that for a single loop one gets 
\begin{equation}  
\Pi^{\mu\nu}(k)=ie^2\int\frac{d^2p}{(2\pi)^2}{\rm  
Tr}\left(\gamma^{\mu}\frac{1}{\not\! p +  
i\varepsilon}\gamma^{\nu}\frac{1}{\not\! p + \not\! k +  
i\varepsilon}\right) = \frac{e^2}{\pi}\left(g^{\mu\nu}- 
\frac{k^{\mu}k^{\nu}}{k^2}\right)\; ,  
\label{vac1}  
\end{equation}  
so that the whole series of Figure~\ref{vac} may be easily  
summed up to give the massive propagator  
\begin{equation}  
D^{\mu\nu}(k) =  \left(-g^{\mu\nu}+  
\frac{k^{\mu}k^{\nu}}{k^2}\right)\frac{1}{\mu^2-k^2}- 
\frac{1}{\lambda}\frac{k^{\mu}k^{\nu}}{(k^2)^2}\; ,  
\label{eq:fo}  
\end{equation} 
with $\mu^2=e^2/\pi$. This is the famous Schwinger boson.

Now we have to consider electron self-energy insertions assuming already
that we have to do with massive photons. This summation is not trivial and
we will perform it in detail. Let us represent the full propagator $S$, in
momentum space, as the sum 
\begin{equation} 
S(p)=\sum_{n=0}^{\infty}S^{(n)}(p)\; , 
\label{sum} 
\end{equation} 

where $S^{(0)}$ is of course the same as Fourier transformed ${\cal
S}_0(x)$ of equation~(\ref{propbet}), and the summation runs over the
number of photons attached to the electron line. To find the recurrent
relation between $S^{(n)}$'s we take the $n$-th term of the
sum~(\ref{sum}) and attach to it the $(n+1)$-st photon. This situation is
schematically represented on Figure~\ref{add}. When the photon is attached
the additional propagator $D^{\mu\nu}(k)$ appears in the internal line. It
may easily be observed that this part of $D^{\mu\nu}$ that bears metric
tensor $g^{\mu\nu}$ does not contribute since the corresponding expression
has the structure
\begin{equation} 
\gamma^{\mu}\gamma^{\alpha_1}\gamma^{\alpha_2}\cdot ...  
\cdot \gamma^{\alpha_{2k+1}}\gamma_{\mu} 
\label{str} 
\end{equation}  

and an odd number of gamma matrices may, in two dimensions, always be
reduced to only one for which one can check that
$\gamma^{\mu}\gamma^{\alpha}\gamma_{\mu} =0$. For the gamma matrices we
use in this work the following convention 
$$   
\gamma^0=\left(\begin{array}{lr}0 & \hspace*{2ex}1 \\ 1 & 0   
\end{array}\right)\; , \;\;\;\;\;   
\gamma^1=\left(\begin{array}{lr} 0 & -1 \\ 1 & 0   
\end{array}\right)\; , \;\;\;\;\;   
\gamma^5=\gamma^0\gamma^1=\left(   
\begin{array}{lr} 1 & 0 \\ 0 & -1 \end{array}\right)\; , 
$$   
and for the metric tensor: $g^{00}=-g^{11}=1$. 
 
Thanks to this observation we may now consider only that part of
$D^{\mu\nu}(k)$ which is proportional to $k^{\mu}k^{\nu}$  
\begin{equation} 
-\frac{k^{\mu}k^{\nu}}{(k^2-\kappa^2)(k^2-\mu^2)}\; ,  
\label{long} 
\end{equation} 
where we have put $\lambda\rightarrow\infty$, and introduced fictious mass
$\kappa^2$ in denominator to avoid infinities at intermediate steps when
we separate the $k$-integral into pieces. Let us now imagine that we first
attach to the object $S^{(n)}(p)$ only one leg of the external photon of
the (incoming) momentum $k$. This means that we consider the vertex in the
$n$-th order: $\left[S(p+k)\Gamma^{\mu}(k,p)S(p)\right]^{(n)}$. But our
(simplified) propagator~(\ref{long}) provides also $k^{\mu}$ in this
vertex.  From the very construction of the theory and its gauge symmetry
it follows that in each order the Ward identity is separately satisfied
which may also be checked by a direct computation 
\begin{equation} 
k_{\mu}\left[S(p+k)\Gamma^{\mu}(k,p)S(p)\right]^{(n)}=
S^{(n)}(p)-S^{(n)}(p+k)\; . 
\label{ward} 
\end{equation} 
If we now attach to the above object the second photon leg (now of
momentum $-k$) we obtain 
\begin{equation} 
\left[S(p-k)\Gamma^{\nu}(-k,p)S(p)- S(p)\Gamma^{\nu}(- 
k,p+k)S(p+k)\right]^{(n)}\; . 
\label{ob} 
\end{equation} 
The second leg, according to~(\ref{long}), also bears  
$k_{\nu}$ so we can use again the Ward identity getting 
\begin{eqnarray} 
k_{\nu}&&\left[S(p-k)\Gamma^{\nu}(-k,p)S(p)-  
S(p)\Gamma^{\nu}(-k,p+k)S(p+k)\right]^{(n)}=\nonumber\\ 
&&=S^{(n)}(p-k) - S^{(n)}(p) + S^{(n)}(p+k) - S^{(n)}(p)\; . 
\label{ward1} 
\end{eqnarray} 
Now we are in a position to state our recurrence equation between $S^{(n)}$'s
\begin{eqnarray} 
&&S^{(n+1)}(p)=\label{rec1}\\ 
&&=-(-
ie)^2\frac{i}{2(n+1)}\int\frac{d^2k}{(2\pi)^2}\frac{1}{(k^2- 
\mu^2+i\varepsilon)(k^2-\kappa^2+i\varepsilon)}\left[2  
S^{(n)}(p-k) - 2 S^{(n)}(p)\right]\; ,\nonumber 
\end{eqnarray}

where in one of the terms in~(\ref{ward1}) we have changed $k\rightarrow
-k$ under the integral. The combinatorical factor $\frac{1}{2(n+1)}$ comes
from the fact that our construction counts each diagram $2(n+1)$ times
($n+1$ possibilities of the choice which photon we treat as the $(n+1)$-st
one and two possibilities of interchanging the attached legs) and $i$ is
required by the Feynman rules ($iD$ on the internal photon line). Finally
we can write this equation in the form 
\begin{equation} 
S^{(n+1)}(p)=\frac{ie^2}{n+1}\left[-{\cal  
I}(\mu^2,\kappa^2)S^{(n)}(p)+\int\frac{d^2k}{(2\pi)^2}\frac{1}
{(k^2-\mu^2+i\varepsilon)(k^2-\kappa^2+i\varepsilon)}S^{(n)}(p-k)\right]\; , 
\label{rec2} 
\end{equation} 
where for convenience symbol $\cal I$ has been introduced to denote 
\begin{equation} 
{\cal I}(\mu^2,\kappa^2)\equiv  
\int\frac{d^2k}{(2\pi)^2}\frac{1}{(k^2-\mu^2+i\varepsilon)(k^2- 
\kappa^2+i\varepsilon)}=\frac{i}{4\pi}\ln\frac{\mu^2/\kappa^ 
2}{\mu^2-\kappa^2}\; . 
\label{ii} 
\end{equation} 
One could observe in this point that the passing to the coordinate space
would simplify further calculations since the convolution integral on the
right hand side of~(\ref{rec2}) would change into product. We decide,
however, to lead all the calculations in momentum space, as one most often
does in field theory, since we find it more instructive.

Repeating the recurrence we are able to write the general formula for the 
$n$-th term 
\begin{eqnarray} 
S^{(n)}(p)=&&\frac{(ie^2)^n}{n!}\sum_{k=0}^{n} 
\frac{n!}{k!(n-k)!}[-{\cal I}(\mu^2,\kappa^2)]^{(n-k)} 
\int\frac{d^2k_1d^2k_2\cdot ... \cdot 
d^2k_k}{(2\pi)^{2k}}\cdot\nonumber\\ 
&&\cdot\frac{1}{(k_1^2-\mu^2+i\varepsilon) (k_1^2- 
\kappa^2+i\varepsilon)}\cdot\frac{1}{(k_2^2- 
\mu^2+i\varepsilon) (k_2^2-\kappa^2+i\varepsilon)}\cdot ...  
\nonumber\\ 
&&\cdot\frac{1}{(k_k^2-\mu^2+i\varepsilon) (k_k^2- 
\kappa^2+i\varepsilon)}\cdot S^{(0)}(p-k_1-k_2- ... -k_k)\; 
.\label{nth} 
\end{eqnarray} 
Considering the summation in~(\ref{sum}) together with that of
formula~(\ref{nth}) we see that the double sum has to be performed. Using
obvious symbolic notation we can simplify it in the following way
\begin{eqnarray}
\sum_{n=0}^{\infty}\frac{x^n}{n!}\sum_{k=0}^{n}\frac{n!}
{k!(n-k)!}a_k &=& \sum_{k=0}^{\infty}\frac{a_k}{k!}
\sum_{n=k}^{\infty}\frac{x^n}{(n-k)!} = 
\sum_{k=0}^{\infty}\frac{a_k x^k}{k!}
\sum_{n=k}^{\infty}\frac{x^{n-k}}{(n-k)!} =\nonumber\\
&=& \sum_{k=0}^{\infty}\frac{a_k x^k}{k!}
\sum_{n=0}^{\infty}\frac{x^n}{n!} = {\rm 
e}^x\sum_{k=0}^{\infty}\frac{a_k x^k}{k!}\; .
\label{tran}
\end{eqnarray}
Applying this to our formula for $S(p)$ we get
\begin{eqnarray} 
S(p) =&& \exp\left[-ie^2{\cal I}(\mu^2,\kappa^2)\right]
\sum_{n=0}^{\infty}\frac{(ie^2)^n}{n!} 
\int\frac{d^2k_1d^2k_2\cdot ... \cdot 
d^2k_n}{(2\pi)^{2n}}\cdot\nonumber\\ 
&&\cdot\frac{1}{(k_1^2-\mu^2+i\varepsilon) (k_1^2- 
\kappa^2+i\varepsilon)}\cdot\frac{1}{(k_2^2- 
\mu^2+i\varepsilon) (k_2^2-\kappa^2+i\varepsilon)}\cdot ...  
\nonumber\\ 
&&\cdot\frac{1}{(k_n^2-\mu^2+i\varepsilon) (k_n^2- 
\kappa^2+i\varepsilon)}\cdot S^{(0)}(p-k_1-k_2- ... -k_n)\; 
.\label{sum1} 
\end{eqnarray} 
We now have to make use of the fact that $S^{(0)}(p)$ is a free massless
propagator: $S^{(0)}(p)=\gamma^{\mu}p_{\mu}/p^2$, pass to the Euclidean
space, and replace denominators $1/D^2$ with $\int_0^{\infty}dt\exp[-t
D^2]$. If we additionally substitute for $p^{\mu}-k_1^{\mu}- ... -
k_n^{\mu}$ the appropriate derivative over $p_{\mu}$ we can write
\begin{eqnarray} 
&&S(p)_E = \frac{1}{2}\exp\left[-ie^2{\cal 
I}(\mu^2,\kappa^2)\right]\left(\gamma_{\mu}\frac{\partial}
{\partial p_{\mu}}\right)_{\! E}\;\;
\sum_{n=0}^{\infty}\frac{(-e^2)^n}{n!}\frac{1}{(\mu^2-
\kappa^2)^n}\int_0^{\infty}\frac{d\tau}{\tau}\int_{0}^{\infty}dt_1
dt_2\cdot ...\cdot dt_n\nonumber\\
&&\cdot\int\frac{d_E^2k_1d_E^2k_2\cdot ... 
\cdot d_E^2k_n}{(2\pi)^{2n}}\sum_{i=0}^{n}\frac{n!}{i!(n-
i)!}(-1)^i\exp\big[-t_1(k_1^2+\mu^2)- t_2(k_2^2+\mu^2)- ...\nonumber\\
&&-t_i(k_i^2+\mu^2)- t_{i+1}(k_{i+1}^2+\kappa^2)- ... - 
t_n(k_n^2+\kappa^2)-\tau (p-k_1-k_2- ... -k_n)^2\big]\; 
.\label{sumeuc} 
\end{eqnarray} 
In this formula the coefficient $1/(\mu^2-\kappa^2)^n$ arises from
expanding the products of denominators $1/[(k_i^2+\mu^2)(k_i^2+\kappa^2)]$
into sums. Now we calculate the multiple integral
\begin{eqnarray}
&&\int\frac{d_E^2k_1d_E^2k_2\cdot ... \cdot 
d_E^2k_n}{(2\pi)^{2n}}\exp\left[-t_1k_1^2-t_2k_2^2- ...
-t_nk_n^2-\tau (p-k_1-k_2- ... -k_n)^2\right]=\nonumber\\
&&=\frac{1}{(4\pi)^n}\frac{1}{\left(1/\tau+1/t_1+ ... 
+1/t_n\right)\tau t_1\cdot ... \cdot t_n}\exp\left[-
p^2/\left(1/\tau+1/t_1+ ... +1/t_n\right)\right]\; .\label{multi}
\end{eqnarray} 
After having taken in~(\ref{sumeuc}) the derivative over $p_{\mu}$ 
the integral over $\tau$ may be easily performed if we observe that
$$
\frac{1}{\left(1/\tau +x\right)^2\tau^2}\exp\left[-\frac{p^2}{1/\tau +x}
\right]=-\frac{1}{p^2}\frac{d}{d\tau}\exp\left[-\frac{p^2}{1/\tau +x}\right]
$$
and one limit contributes $\frac{1}{p^2}$ and the other
$-\frac{1}{p^2}{\rm e}^{-p^2/x}$. In that way we obtain for $S(p)_E$
\begin{eqnarray} 
&&S(p)_E =\exp\left[-ie^2{\cal I}(\mu^2,\kappa^2)\right]\frac{
\left(\gamma_{\mu}p^{\mu}\right)_E}{p^2}
\sum_{n=0}^{\infty}\frac{1}{n!}\left(\frac{-e^2}{4\pi(\mu^2-
\kappa^2)}\right)^n \int_0^{\infty}dt_1dt_2\cdot ...\cdot 
dt_n\cdot\nonumber\\
&&\exp\left[-\kappa^2(t_1+t_2+ ... +t_n)\right]\left(\exp\left[-
\frac{p^2}{1/t_1+1/t_2+ ... +1/t_n}\right]-
1\right)\sum_{i=0}^{n}\frac{n!}{i!(n-i)!}(-1)^i\cdot\nonumber\\
&&\cdot\exp\left[-
(\mu^2-\kappa^2)(t_1+t_2+ ... +t_i)\right]\; .\label{sue} 
\end{eqnarray} 
Now let us consider the expression under the second sum 
$$
\sum_{i=0}^{n}\frac{n!}{i!(n-i)!}(-1)^i f(t_1)f(t_2)\cdot ... 
\cdot f(t_i)\; .
$$
Since it will be integrated in~(\ref{sue}) over all $t_i$'s  with 
a symmetric function of its arguments one can obviously replace it with
$$
[1-f(t_1)]\cdot[1-f(t_2)]\cdot ... \cdot[1-f(t_n)]
$$
and that, in turn, leads to
\begin{eqnarray} 
S(p)_E =&&\exp\left[-ie^2{\cal 
I}(\mu^2,\kappa^2)\right]\frac{\left(\gamma_{\mu}p^{\mu}
\right)_E}{p^2}
\sum_{n=0}^{\infty}\frac{1}{n!}\left(\frac{-e^2}{4\pi(\mu^2-
\kappa^2)}\right)^n \int_0^{\infty}dt_1dt_2\cdot ...\cdot 
dt_n\label{sue2}\\
&&\frac{{\rm e}^{-\kappa^2 t_1}-{\rm e}^{-\mu^2 
t_1}}{t_1}\cdot\frac{{\rm e}^{-\kappa^2 t_2}-{\rm e}^{-
\mu^2 t_2}}{t_2}\cdot ... \cdot\frac{{\rm e}^{-\kappa^2 t_n}-
{\rm e}^{-\mu^2 t_n}}{t_n}\cdot\nonumber\\
&&\cdot\left(\exp\left[-
\frac{p^2}{1/t_1+1/t_2+ ... +1/t_n}\right]-1\right)\; 
.\nonumber 
\end{eqnarray} 
Making now use of the identity which is valid for $a>0$~\cite{gr}
$$
1-{\rm e}^{-1/4a}=\int_0^{\infty}dx J_1(x){\rm e}^{-a x^2}\; ,
$$
where $J_1$ is the Bessel function, together with the substitution: 
$$
\frac{1}{4a}=\frac{p^2}{\left(\frac{1}{t_1}+\frac{1}{t_2}+... 
+\frac{1}{t_1}\right)}\; ,
$$ 
we note that we are now in a position to perform all $t_i$ integrations
according to
$$
\int_0^{\infty}dt \frac{{\rm e}^{-\kappa^2 t}-{\rm e}^{-\mu^2 t}}{t}
{\rm e}^{-x^2/4p^2t}=2\left[K_0\left(\frac{\kappa x}{\sqrt{p^2}}\right)
- K_0\left(\frac{\mu x}{\sqrt{p^2}}\right)\right]\; .
$$
Identifying that in~(\ref{sue2}) we have in fact the expansion of the
exponent function we can write down the following formula
\begin{eqnarray} 
S(p)_E =&&-\exp\left[-ie^2{\cal 
I}(\mu^2,\kappa^2)\right]\frac{\left(\gamma_{\mu}p^{\mu}
\right)_E}{p^2}\int_0^{\infty}dx J_1(x)\nonumber\\
&&\exp\left\{-\frac{e^2}{2\pi(\mu^2-\kappa^2)} 
\left[K_0\left(\frac{\kappa x}{\sqrt{p^2}}\right)- 
K_0\left(\frac{\mu x}{\sqrt{p^2}}\right)\right]\right\}\; 
.\label{sue3} 
\end{eqnarray} 
The quantity $\kappa$ was introduced to the calculations only temporarily
in order to regularize certain integrals on intermediate steps.  Now, in
the formula~(\ref{sue3}), where all pieces are collected together, we may
get rid of it, setting $\kappa\rightarrow 0$, if we make use of the
expansion of Basset function for small arguments: $K_0(x)\approx
-\ln(x/2)-\gamma_E$. Recalling that $\mu^2=e^2/\pi$ we finally get
\begin{equation}
S(p)_E=-\frac{(\not\! p)_E}{(p^2)^{5/4}}{\rm 
e}^{\gamma_E/2}\left(\frac{e}{2\sqrt{\pi}}\right)^{1/2}\int_0
^{\infty}dx x^{1/2}J_1(x)\exp\left[\frac{1}{2}K_0\left(e 
x/\sqrt{\pi p^2}\right)\right]\; .
\label{fina}
\end{equation}
If one takes into account the asymptotic approximation of the function $K_0$
one can easily obtain the known~\cite{stam} infrared behaviour of the
electron propagator (in Minkowski space):
$S(p)\approx\frac{e^{1/2}}{2^{5/2}\pi^{5/4}}\exp\left( 
\frac{\gamma_E}{2}\right)\left[\Gamma\left(\frac{1}{4}\right)\right]^2
\frac{\not p}{(-p^2)^{5/4}}$.

Then we already have the Euclidean $p$ representation of $S$, but what we
need in order to compare the result with~(\ref{propbet}) and~(\ref{beta})
is the coordinate space representation. The lacking Fourier transform may,
however, be performed in a straightforward way described below. After
rescaling $x\rightarrow x\cdot p$, where $p=\sqrt{p^2}$, replacing
$p^{\mu}/p$ with $\partial/\partial p_{\mu}$ and noticing that
$J_1(x)=-dJ_0(x)/dx$ one gets
\begin{equation}
S(p)_E=\left(\gamma_{\mu}\frac{\partial}{\partial p_{\mu}}\right)_E {\rm 
e}^{\gamma_E/2}\left(\frac{e}{2\sqrt{\pi}}\right)^{1/2}\int_0
^{\infty}dx x^{-1/2}J_0(x p)\exp\left[\frac{1}{2}K_0\left(e 
x/\sqrt{\pi}\right)\right]\; .
\label{four1}
\end{equation}
The following representation for the Bessel function $J_0(x)$
$$
J_0(x p) =\frac{1}{2\pi}\int_0^{2\pi}{\rm 
e}^{ipx\sin(\phi-\alpha)}d\phi
$$
can now be used, where we choose the angle $\alpha$ such that: $\cos\alpha
= p_4/p$, and $\sin\alpha=p_1/p$. After this substitution our
formula~(\ref{four1}) contains two integrations: over $x$ and $\phi$ and
they may be replaced with the integration over Euclidean two-space if we
identify
$$
x_4=x\sin\phi\; ,\;\;\;\;\; x_1=x\cos\phi\; .
$$
Taking into account that the appropriate Jacobian equals $1/x$ and passing
to Minkowski space-time we can finally write down
\begin{equation}
S(p)=-\frac{1}{2\pi}e^{\gamma_E/2}\int d^2x {\rm e}^{ipx}
\frac{\not\! x}{x^2-i\varepsilon}\exp\left[\frac{1}{2}\ln\sqrt{-
e^2x^2/4\pi}+\frac{1}{2}K_0\left(\sqrt{-e^2x^2/\pi}\right)
\right]\; ,
\label{end}
\end{equation}
which entirely agrees with the formulae~(\ref{propbet}) and (\ref{beta})
in the case when $x$ is spacelike. For timelike $x$ we have to perform in
~(\ref{fina}) a rotation in the complex plane of $p^2$ obtaining the first
formula of~(\ref{beta}).  This proves the convergence and correctness of the
perturbation series in the Schwinger Model (at least for the electron
propagator).

\acknowledgements
The author would like to thank to Professor J. Namys{\l}owski 
for the interesting discussions.

\setlength{\unitlength}{0.012pt}  
\begin{figure}[p]   
\begin{center}   
\begin{picture}(31000,14000) 
\thicklines
\put(4300,6950){\oval(600,600)[t]}  
\put(4900,7050){\oval(600,600)[b]}  
\put(5500,6950){\oval(600,600)[t]}  
\put(6100,7050){\oval(600,600)[b]}  
\put(6700,6950){\oval(600,600)[t]}  
\put(7300,7050){\oval(600,600)[b]}  
\put(7900,6950){\oval(600,600)[t]}  
\put(8500,7050){\oval(600,600)[b]}  
\thinlines

\put(9900,6950){\large =}  

\put(12200,6950){\oval(600,600)[t]}  
\put(12800,7050){\oval(600,600)[b]}  
\put(13400,6950){\oval(600,600)[t]}  
\put(14000,7050){\oval(600,600)[b]}  
\put(14600,6950){\oval(600,600)[t]}  
\put(15200,7050){\oval(600,600)[b]}  
\put(15800,6950){\oval(600,600)[t]}  
\put(16400,7050){\oval(600,600)[b]}  

\put(17800,6950){\large +}  

\put(20100,6950){\oval(600,600)[t]}  
\put(20700,7050){\oval(600,600)[b]}  
\put(21300,6950){\oval(600,600)[t]}  
\put(21900,7050){\oval(600,600)[b]}  
\put(22900,7000){\circle{1400}}   
\put(23900,6950){\oval(600,600)[t]}  
\put(24500,7050){\oval(600,600)[b]}  
\put(25100,6950){\oval(600,600)[t]}  
\put(25700,7050){\oval(600,600)[b]}  

\put(27100,6950){\large +}  
  
\put(300,1950){\oval(600,600)[t]}  
\put(900,2050){\oval(600,600)[b]}  
\put(1500,1950){\oval(600,600)[t]}  
\put(2100,2050){\oval(600,600)[b]}  
\put(3100,2000){\circle{1400}}   
\put(4100,1950){\oval(600,600)[t]}  
\put(4700,2050){\oval(600,600)[b]}  
\put(5300,1950){\oval(600,600)[t]}  
\put(5900,2050){\oval(600,600)[b]}  
\put(6900,2000){\circle{1400}}   
\put(7900,1950){\oval(600,600)[t]}  
\put(8500,2050){\oval(600,600)[b]}  
\put(9100,1950){\oval(600,600)[t]}  
\put(9700,2050){\oval(600,600)[b]}  

\put(11100,1950){\large +}  

\put(13400,1950){\oval(600,600)[t]}  
\put(14000,2050){\oval(600,600)[b]}  
\put(14600,1950){\oval(600,600)[t]}  
\put(15200,2050){\oval(600,600)[b]}  
\put(16200,2000){\circle{1400}}   
\put(17200,1950){\oval(600,600)[t]}  
\put(17800,2050){\oval(600,600)[b]}  
\put(18400,1950){\oval(600,600)[t]}  
\put(19000,2050){\oval(600,600)[b]}  
\put(20000,2000){\circle{1400}}   
\put(21000,1950){\oval(600,600)[t]}  
\put(21600,2050){\oval(600,600)[b]}  
\put(22200,1950){\oval(600,600)[t]}  
\put(22800,2050){\oval(600,600)[b]}  
\put(23800,2000){\circle{1400}}   
\put(24800,1950){\oval(600,600)[t]}  
\put(25400,2050){\oval(600,600)[b]}  
\put(26000,1950){\oval(600,600)[t]}  
\put(26600,2050){\oval(600,600)[b]}  

\put(28000,1950){\large + \hskip 2mm ...}  
\end{picture}  
\caption{Diagrams contributing to the photon  
propagator.}  
\label{vac}  
\end{center}  
\end{figure}
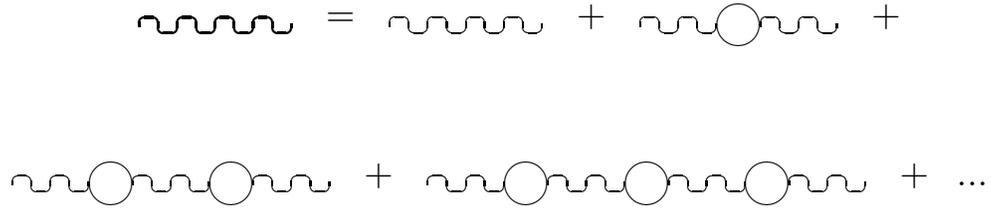   
  
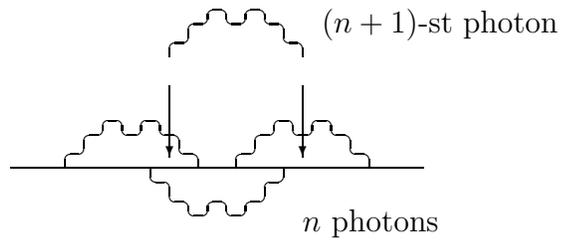
\begin{figure}[p]   
\begin{center}   
\begin{picture}(15000,10000)  
\put(0,2000){\line(1,0){13000}}  
\put(2000,2000){\oval(600,900)[tl]}  
\put(2000,2750){\oval(600,600)[br]}  
\put(2600,2750){\oval(600,600)[tl]}  
\put(2600,3350){\oval(600,600)[br]}  
\put(3200,3200){\oval(600,600)[t]}  
\put(3800,3350){\oval(600,600)[b]}  
\put(4400,3200){\oval(600,600)[t]}  
\put(5000,3350){\oval(600,600)[bl]}  
\put(5000,2750){\oval(600,600)[tr]}  
\put(5600,2750){\oval(600,600)[bl]}  
\put(5600,2000){\oval(600,900)[tr]}  
  
\put(7400,2000){\oval(600,900)[tl]}  
\put(7400,2750){\oval(600,600)[br]}  
\put(8000,2750){\oval(600,600)[tl]}  
\put(8000,3350){\oval(600,600)[br]}  
\put(8600,3200){\oval(600,600)[t]}  
\put(9200,3350){\oval(600,600)[b]}  
\put(9800,3200){\oval(600,600)[t]}  
\put(10400,3350){\oval(600,600)[bl]}  
\put(10400,2750){\oval(600,600)[tr]}  
\put(11000,2750){\oval(600,600)[bl]}  
\put(11000,2000){\oval(600,900)[tr]}  
  
\put(4700,2000){\oval(600,900)[bl]}  
\put(4700,1250){\oval(600,600)[tr]}  
\put(5300,1250){\oval(600,600)[bl]}  
\put(5300,650){\oval(600,600)[tr]}  
\put(5900,800){\oval(600,600)[b]}  
\put(6500,650){\oval(600,600)[t]}  
\put(7100,800){\oval(600,600)[b]}  
\put(7700,650){\oval(600,600)[tl]}  
\put(7700,1250){\oval(600,600)[br]}  
\put(8300,1250){\oval(600,600)[tl]}  
\put(8300,2000){\oval(600,900)[br]}  
\put(9200,0){$n$ photons}  
  
\put(5300,5500){\oval(600,900)[tl]}  
\put(5300,6250){\oval(600,600)[br]}  
\put(5900,6250){\oval(600,600)[tl]}  
\put(5900,6850){\oval(600,600)[br]}  
\put(6500,6700){\oval(600,600)[t]}  
\put(7100,6850){\oval(600,600)[b]}  
\put(7700,6700){\oval(600,600)[t]}  
\put(8300,6850){\oval(600,600)[bl]}  
\put(8300,6250){\oval(600,600)[tr]}  
\put(8900,6250){\oval(600,600)[bl]}  
\put(8900,5500){\oval(600,900)[tr]}  
\put(5000,4600){\vector(0,-1){2250}}  
\put(9200,4600){\vector(0,-1){2250}}  
\put(9800,6300){$(n+1)$-st photon}   
  
\end{picture}  
\end{center}  
\caption{The attachment of the $(n+1)$-st photon to  
$S^{(n)}(p)$.}  
\label{add}  
\end{figure}  
\end{document}